\documentclass[final, conference]{IEEEtran}
\IEEEpubid{}
\IEEEoverridecommandlockouts
\usepackage{cite}
\usepackage{amsmath,amssymb,amsfonts}
\usepackage{algorithmic}
\usepackage{graphicx}
\usepackage{textcomp}
\usepackage{xcolor}
\def\BibTeX{{\rm B\kern-.05em{\sc i\kern-.025em b}\kern-.08em
    T\kern-.1667em\lower.7ex\hbox{E}\kern-.125emX}}

\usepackage{todonotes}

\usepackage{enumitem}
\setenumerate{itemsep=2pt,topsep=2pt,leftmargin=17pt,parsep=0pt,partopsep=0pt}
\setitemize{itemsep=2pt,topsep=2pt,leftmargin=17pt,parsep=0pt,partopsep=0pt}

\DeclareMathOperator*{\argmax}{arg\,max}

\usepackage{hyperref}

\usepackage{float}

\begin{document}
\title{Weaponizing Unicodes with Deep Learning - Identifying Homoglyphs with Weakly Labeled Data\\
}

\author{Anonymous}

\author{
\IEEEauthorblockN{Perry Deng\textsuperscript{\textsection}}
\IEEEauthorblockA{\textit{Global Cybersecurity Institute} \\
\textit{Rochester Institute of Technology}\\
Rochester, USA \\
perry.deng@mail.rit.edu}
\and
\IEEEauthorblockN{Cooper Linsky\textsuperscript{\textsection}}
\IEEEauthorblockA{\textit{Global Cybersecurity Institute} \\
\textit{Rochester Institute of Technology}\\
Rochester, USA \\
cwl4602@rit.edu}
\and
\IEEEauthorblockN{Matthew Wright}
\IEEEauthorblockA{\textit{Global Cybersecurity Institute} \\
\textit{Rochester Institute of Technology}\\
Rochester, USA \\
matthew.wright@rit.edu}
}

\maketitle

\newcommand\nnfootnote[1]{%
  \begin{NoHyper}
  \renewcommand\thefootnote{}\footnote{#1}%
  \addtocounter{footnote}{-1}%
  \end{NoHyper}
}
\nnfootnote{This material is based upon work supported by the National Science Foundation under Award No. 1816851.}
\nnfootnote{\textsuperscript{\textsection}These authors contributed equally to this work.}
\nnfootnote{\copyright{}2020 IEEE}

\begin{abstract}
    Visually similar characters, or \emph{homoglyphs}, can be used to perform social engineering attacks or to evade spam and plagiarism detectors. It is thus important to understand the capabilities of an attacker to identify homoglyphs -- particularly ones that have not been previously spotted -- and leverage them in attacks. We investigate a deep-learning model using embedding learning, transfer learning, and augmentation to determine the visual similarity of characters and thereby identify potential homoglyphs. Our approach uniquely takes advantage of weak labels that arise from the fact that most characters are not homoglyphs. Our model drastically outperforms the Normalized Compression Distance approach on pairwise homoglyph identification, for which we achieve an average precision of 0.97. We also present the first attempt at clustering homoglyphs into sets of equivalence classes, which is more efficient than pairwise information for security practitioners to quickly lookup homoglyphs or to normalize confusable string encodings. To measure clustering performance, we propose a metric (mBIOU) building on the classic Intersection-Over-Union (IOU) metric. Our clustering method achieves 0.592 mBIOU, compared to 0.430 for the naive baseline. We also use our model to predict over 8,000 previously unknown homoglyphs, and find good early indications that many of these may be true positives. Source code and list of predicted homoglyphs are uploaded to Github: https://github.com/PerryXDeng/weaponizing\_unicode
\end{abstract}

\section{Introduction}
\label{intro}
\begin{figure}[t]
\centerline{\includegraphics[width=.45\textwidth]{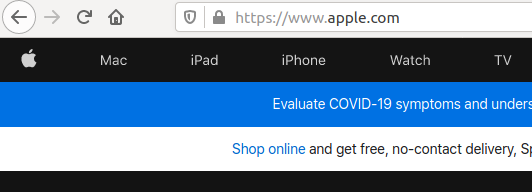}}
\vskip -0.2cm
\caption{A screenshot of \url{apple.com}, which uses the Latin script}
\label{fig:apple}
\end{figure}

\begin{figure}[t]
\centerline{\includegraphics[width=.45\textwidth]{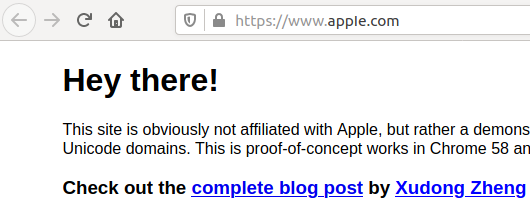}}
\vskip -0.2cm
\caption{A screenshot of \cite{zheng_2017}'s domain, which uses the Cyrillic script}
\label{fig:apple_fake}
\end{figure}

Identical-looking characters, known as homoglyphs, can be used to substitute characters in strings to create visually similar strings. These strings can then be used to trick users into clicking on malicious domains passing as legitimate ones~\cite{gabrilovich2002homograph}, or to evade spam and plagiarism detectors~\cite{weber2013plagiarism}. While the security risks of homoglyphs have been known since the turn of the century~\cite{gabrilovich2002homograph}, the problem has gotten more attention as Unicode becomes the predominant standard for text processing. Unlike ASCII, which contains only 128 characters, the latest Unicode standard encapsulates more than 140 \emph{thousand} characters. Among the Unicode characters, many have been identified to be homoglyphs. This allows scenarios where the entire string of a domain name can be substituted with different Unicode characters and appear visually identical (See Figures~\ref{fig:apple} and \ref{fig:apple_fake}). A complete database of all homoglyphs within the Unicode standard would be helpful for us to assess the risks of systems utilizing Unicode. Due to the sheer size of the Unicode standard, however, manual identification of all homoglyphs is infeasible. 


Prior works~\cite{roshanbin_miller_2011, ginsberg2018rapid, sawabe2019detection} on automated homoglyph identification have a few critical shortcomings: 1) they are only used to identify homoglyphs for few characters, or/and 2) they do not cluster homoglyphs into distinct sets, or/and 3) they do not present quantifiable evidence on the accuracy of their approach. We seek to address these shortcomings with empirical evidence on a method based on deep learning, which has performed tremendously well on computer vision problems such as object classification~\cite{tan2019efficientnet} and medical imaging~\cite{pisano2020ai}. Unlike many works in computer vision, we do not have a large number of human-labeled examples for each visual class of data, because we do not know the complete sets of homoglyphs. We only know that the overwhelming majority of characters are \emph{not} homoglyphs.

To address this challenge, we propose a novel model that adapts data augmentation, transfer learning, and embedding learning from similarly challenging deep-learning problems such as facial recognition. We fine-tune our neural network on weakly labeled data by exploiting the fact that, in general, different characters do not look alike. We compare our deep learning model to the approach of Roshabin and Miller~\cite{roshanbin_miller_2011} on the pairwise identification of homoglyphs. We show that our method significantly outperforms theirs in both accuracy (0.91 vs 0.64) and average precision (0.97 vs 0.75). We also attempt clustering of homoglyphs into distinct sets, and propose to measure the performance and generalizability of our clustering model with mBIOU, a metric based on Intersection Over Union. Our approach achieves a mBIOU of 0.592 between the predicted sets of homoglyphs and the known homoglyphs from Unicode Consortium~\cite{uts39}. In addition, we show that our learned deep embeddings can be used to find many previously unidentified homoglyphs.

In summary, our contributions are as follows: 
\begin{itemize}
    \item We propose a novel deep learning model for identifying homoglyphs that overcomes the lack of training data by leveraging embedding learning and transfer learning, together with data augmentation methods.
    \item We are the first to attempt clustering of homoglyphs into distinct sets, and propose mBIOU, a way to systematically measure the accuracy of clustered sets.
    \item We show that our approach outperforms the prior work on the pairwise identification task, with an average precision of 0.97 compared to 0.79. Our approach also achieves 0.592 mBIOU on the clustering task, compared to a naive baseline of 0.430.
    \item We apply our model to find new homoglyphs, and visualize a random sample of predicted sets. 8,452 unrecorded homoglyphs were predicted with our approach, and visual inspection of our random sample suggests that some of these are likely to be true positives.
\end{itemize}

\section{Background}
\label{background}

\subsection{Unicode}
\label{background:unicode}
Unicode~\cite{unicode} is the predominant technology standard for the encoding, representation, and processing of text. It is maintained by the Unicode Consortium, and is backward-compatible with the legacy ASCII.
Unicode defines a space of integers, or \emph{codepoints}, ranging from 0 to 0x10FFFF, which is hexadecimal for slightly more than one million. 
The entire set of codepoints is called the \emph{code space}. The codepoints can be mapped to \emph{characters}, so that characters can be encoded and processed as unique numbers by computers. 
As of Unicode Version 13.0 released in March 2020, there are 143,859 characters mapped to codepoints in the standard implemented for public use. 
For human readability, a codepoint can either be rendered into an image with a font, or printed in hex format with "U+\textless hex code\textgreater ". For example, the lowercase first letter of the Latin alphabet can be 'a' or U+0061.

\subsection{Homoglyphs, Homographs, and Confusables}
A \emph{homoglyph} is a character that is visually similar to another character. For example, U+0049, or 'I', can appear similar to U+006C, or 'l'. 
Homoglyphs can be used to substitute characters in strings of one or more characters to form \emph{homographs},
or \emph{confusables} in Unicode terminology. Confusables pose significant security risks. They can fool users into unintended actions, such as malicious domains masquerading as legitimate websites~\cite{gabrilovich2002homograph}. Confusables can also cause text to slip past mechanical gatekeepers, such as a spam or plagiarism detector~\cite{weber2013plagiarism}.

The Unicode Consortium maintains a database~\cite{uts39} of confusables to help mitigate security risks. The incomplete database maps visually similar strings of one or more codepoints to \emph{equivalence classes}, identified by their respective ``prototype" string. With this mapping, security practitioners can quickly lookup homoglyphs for a given character or normalize confusable strings into a single encoding~\cite{uts39}. Formally, given the set of possible Unicode strings $U^*$, and the equivalence relation "is visually similar to", $\equiv$, on $U^*$, the confusable equivalence class $S$ of a string $\vec{u}$ is: 
\begin{gather}
\label{eq:confusables_equivalence}
S(\vec{u}) = \{\vec{s} \in U^* | \vec{s} \equiv \vec{u}, \vec{u} \in U^*\}
\end{gather}
Since there are infinite strings, developing an exhaustive confusable database is impossible. We can instead limit the scope to just single codepoints. More formally, we can modify Equation~\ref{eq:confusables_equivalence} to have equivalence class (or simply \emph{class}) $H$ of a codepoint $c$:
\begin{gather}
\label{eq:homoglyphs_equivalence}
H(c) = \{h \in U | h \equiv c, c \in U\}
\end{gather}
where $U$ is the set of all renderable Unicode codepoints. When $|H(c)| \geq 2$, we get \emph{non-trivial sets} of two or more homoglyphs (as each character is always in an equivalence class containing at least itself). With these homoglyph sets, we can form longer confusables through character substitution and assess risks in vulnerable systems. 

\subsection{Automated Homoglyph Identification}
\label{background:identification}
Homoglyph identification can be understood as the problem of finding all sets of equivalence classes that satisfy Equation~\ref{eq:homoglyphs_equivalence}. An alternative problem formation, used in all previous work, is to identify pairs of homoglyphs rather than sets. Pairs are less useful than sets, because for us to store all the homoglyph associations of all given characters, pairwise lookup would require drastically more storage. To the best of our knowledge, prior work on the automated identification of homoglyphs within Unicode includes:
\begin{itemize}
    \item Roshanbin and Miller~\cite{roshanbin_miller_2011} develop an approach based on Normalized Compression Distance (NCD) 
    that identifies \emph{pairs} of homoglyphs from a set of around 6,200 Unicode characters. They claim that their study expands the set of known homoglyphs at the time of their work.
    
    \item Ginsberg and Yu~\cite{ginsberg2018rapid} aim to identify potential homoglyph pairs by decreasing their granularity and normalizing their sizes and locations. They present empirical results on a select few letters, and mainly focus on the speed of their method. Like Roshanbin and Miller, they did not attempt to map codepoints into equivalence sets.
    
    \item Sawabe et al.~\cite{sawabe2019detection} use optical-character-recognition (OCR) software to produce a dynamic mapping from Unicode characters to ASCII characters. They assesses the OCR homoglyph detection scheme as part of a larger model, evaluating its utility to detect confusable URLs against existing Unicode-ASCII mappings. Although OCR technologies have become more sophisticated and effective in recent years, they are typically designed to recognize a small subset of Unicode. This makes it impossible to identify more equivalence classes than the number of visually distinct characters supported by the OCR software, which is fewer than 128.
\end{itemize}
No prior works quantify the accuracy of their approach.

\subsection{Few-Shot Learning}
\label{background:fewshot}
Deep-learning approaches typically perform poorly without a lot of data. Recent work has attempted to overcome this by addressing the so-called \emph{few-shot} learning problem, where just a few examples are given for a class.
Few-Shot classification or clustering problems can be described as "$i$-shot, $j$-way", where $i$ stands for the number of examples given for a class, and $j$ is the number of classes. Many approaches~\cite{wang2020generalizing} have been proposed to improve the performance of deep-learning methods in few-shot learning, with the most promising ones being data augmentation, embedding learning, and multitask/transfer learning. We leverage some of these techniques in our work.

\section{Deep Learning Approach}
\label{approach}

\subsection{Learning an Embedding Function}
\label{training}
Classifying visually similar glyphs into distinct sets can be considered as a computer vision task. There has been prior work in computer vision tackling similar problems, such as using Convolutional Neural Networks (CNNs) for 1-shot, 20-way classification on the Omniglot dataset~\cite{Lake1332}. The Omniglot dataset is similar to our problem, in that the problem is glyph classification. It contains 1623 distinct characters, and during testing, the model is asked to predict between 20 character classes. There are, however, many more than 20 equivalence classes (Equation~\ref{eq:homoglyphs_equivalence}) in Unicode. For instance, if (conservatively) 80\% of codepoints do not look similar, then we would have more than 115,000 classes. In that aspect, our problem is more like mass scale facial recognition than 20-way character classification. 

To tackle this, we utilize several techniques from the few-shot learning problem domain, including data augmentation, transfer learning, and embedding learning. 

\subsubsection*{Data Augmentation}
One issue with our data is not knowing the equivalence class membership of a character in general. In other words, we lack labels for nearly all characters. While there are public databases of known homoglyphs, there is no guarantee that they are representative of the entire codespace. The confusable database maintained by Unicode Consortium~\cite{uts39}, for example, underrepresents Chinese-Japanese-Korean characters in their database, leading us to suspect that most homoglyphs are still unrecorded. In addition, training on the few known class labels makes it harder to assess the model's ability to generalize. We therefore forego the use of known homoglyph labels for training. Instead, we generate \emph{weak labels} by exploiting the fact that, in general, characters are not homoglyphs, and belong to their own trivial equivalence classes. This allows us to sample characters from different classes by random uniform choice, with reasonable reliability. 

Another issue is that we only have a few example images for each codepoint, from various fonts' implementations of the same character. This means that we only have a few datapoints for each weakly labeled class. To alleviate this problem, we augment our data by randomizing the location of a glyph within an image and performing random affine transformations on images during training.

\subsubsection*{Embedding Learning}
Embedding learning is a type of modeling where, instead of directly learning a classification or regression model,
we learn an \emph{embedding function} that outputs relevant latent variables. Embedding learning excels in unsupervised and semi-supervised few-shot learning problems~\cite{wang2020generalizing}. Neural network architectures are good candidates for the embedding functions, because they can approximate highly non-linear functions. 

To train our embedding function, we utilize a modified version of the Triplet Loss function from Facenet~\cite{schroff2015facenet}, a work on facial recognition. In Triplet Loss, a triplet is consisted of an \emph{anchor} datapoint, a \emph{positive} datapoint from the same class as the anchor, and a \emph{negative} datapoint from a different class. Having computed the embeddings for all three samples, Triple Loss is optimized when the distance between the anchor and positive embeddings is low compared to the distance between the anchor and negative embeddings.

To produce our triplets, we use the weak labels to generate an anchor and a positive sample from the same codepoint, with a negative sample from a different codepoint. For computing Triplet Loss, though a Euclidean distance between embeddings might be used, we instead use one minus the cosine similarity.
We also did not implement the complicated triplet mining algorithm proposed by Schroff et al.~\cite{schroff2015facenet}. 

\subsubsection*{Transfer Learning}
For our embedding function, we use EfficientNet~\cite{tan2019efficientnet} without its softmax classification layer. It is a CNN, and the feature maps of its penultimate layer are flattened to become our embedding vectors. EfficientNet is pretrained on ImageNet, and has achieved state-of-the-art transfer performance in various datasets~\cite{tan2019efficientnet}. We apply a regularization loss by penalizing the $L_2$ distance between our learned weights and the pretrained weights.

\subsection{Clustering Homoglyphs into Equivalence Classes}
\label{cluster}
Given the learned embedding function, we can separate the embeddings into equivalence classes by assigning them to distinct clusters. While many methods exist for unsupervised clustering problems, most are designed with a fixed number of clusters in mind. They are thus not suited for our problem, where a huge unknown number of single datapoint clusters (non-homoglyph characters) exist. We propose an ad-hoc heuristic, where each character is iterated through, assigned to an existing cluster where all characters have a similarity with the candidate character greater than a particular threshold, or a new cluster of its own. Afterwards, we iterate through all proposed clusters $C_i$ in a list, comparing them to the merge candidates, $C_{k, k > i}$, that are clusters further down the list. We calculate the mean and variance of cosine similarities between the characters in each $C_k$ and characters in $C_i$, and merge $C_k$ with $C_i$ if certain thresholds apply to the mean and variance of the cosine similarities.

\subsection{Rendering the Codepoints}
\label{rendering}
To render $U$, we use a collection of Google NoTo fonts, and Windows and MacOS default fonts. For consistency, we convert each font to the TrueType format, the most common format in the set. Each TrueType font contains a list of supported Unicode codepoints and a graphical representation of how they are displayed. Although there are 143,859 characters mapped to codepoints by the Unicode Consortium, a codepoint can only be utilized if it is renderable by at least one font. For a codepoint to be renderable by a font, it must meet the following criteria:
\begin{itemize}
    \item The character must be on the font’s list of supported characters.
    \item The font does not render the character as a replacement character, which occurs when a font is unable to represent a Unicode character graphically. 
\end{itemize}
116,294 characters meet these criteria and are used for training and testing. During training, we treat multiple font implementations of the same character as multiple datapoints of the same weakly labeled class, and the specific font to render a character is selected at random. In our experiments for model assessment and comparison, we follow prior work and treat each character as a single datapoint. As Roshanbin and Miller point out~\cite{roshanbin_miller_2011}, ideally, the same font would render every codepoint. However, there is no such font that supports rendering every Unicode character. We therefore use the minimum possible number of fonts, as~\cite{roshanbin_miller_2011} does, in our experiments.

\section{Experiments}
\label{experiments}
\subsection{Metrics: Clustering}
\label{methodology}

To assess the efficacy of our approach at clustering Unicode codepoints into homoglyph equivalence classes, we conduct several experiments on our trained model to show our approach predicts equivalence classes that are close to ground truth. Since the ground truth equivalence classes for Equation~\ref{eq:homoglyphs_equivalence} are unknown over the set $U$, it is impossible for us to directly measure the fit of predicted equivalence classes. Luckily, Unicode Consortium~\cite{uts39} maintains a database of known confusable strings. About 5000 of these confusable strings are length one strings, or homoglyph characters, which are useful to compare our predictions against. These homoglyphs are grouped into distinct sets, which satisfy our definition of equivalence classes. 4666 of these homoglyphs can be rendered by the fonts built into our operating systems. This gives us a total of 1313 non-trivial equivalence classes. For the rest of this section, we refer to these as the ground truth equivalence classes on $U'$. Formally, let's define $U'$ as the set of 4666 known homoglyphs from Unicode Consortium, we can then substitute $U$ with $U'$ in Equation \ref{eq:homoglyphs_equivalence}, which yields 1313 ground truth equivalence classes satisfying the definition:
\begin{gather}
\label{eq:consortium_equivalence}
H'(c) = \{h \in U' | h \equiv c, c \in U'\}
\end{gather}
Most of the 1313 equivalence classes have a cardinality no more than 10, with the largest being 71 (Table \ref{table:class_sizes}). We refer to these ground truth individual equivalence classes as $H'_{1}, H'_{2}, \dots, H'_{1313}$. Likewise, we refer to our each of our $k$ predicted equivalence classes as 
$H^p_{1}, H^p_{2}, \dots, H^p_{k}$,
where $k$ can be any number of equivalence classes predicted, and $p$ simply indicates "predicted."

\begin{table}[htbp]
\renewcommand{\arraystretch}{1.2}
\caption{Distribution of Equivalence Class Cardinality in $U'$}
\begin{center}
\begin{tabular}{cc}
Cardinality & Number of Equivalence Classes \\
\hline
2 - 10 & 1237 \\
11 - 20 & 45 \\
21 - 30 & 23 \\
31 - 40 & 5 \\
41 - 50 & 1 \\
51 - 60 & 0 \\
61 - 70 & 1 \\
71 - 80 & 1 \\
\end{tabular}
\label{table:class_sizes}
\end{center}
\vspace{-10pt}
\end{table}

Therefore, we estimate the fit by measuring the distance between our equivalence class predictions (predicted sets) and the ground truth sets over $U'$ rather than $U$. The metric we use is \emph{mean Best Intersection Over Union} (mBIOU). Intersection Over Union is a way to measure the distance between two sets, commonly used in object detection problems. mBIOU first finds the best matching between the predicted sets and ground truth sets. For example, if there is a predicted set of characters that look like the letter 'a', it will be matched with a similar set of ground truth characters, if such a set exists. Given this matching, mBIOU computes the average IOU between the matched pairs. 


Formally, we define 
\begin{gather}
\label{eq:miou}
    mBIOU = \frac{1}{n}\times\sum^{n}_{i=1} \frac{| H'_{i} \cap H^{best}_{i} |}{| H'_{i} \cup H^{best}_{i} |}
\end{gather}
where $n = 1313$ is the total number of ground truth equivalence classes in $U'$, and 
\begin{equation}
\begin{gathered}
\label{eq:best_predicted_class}
    H^{best}_{i} = \argmax_{H^p_{j}} | H'_{i} \cap H^p_{j} |  \\
    \textrm{s.t.} \quad 1 \leq j \leq k
\end{gathered}
\end{equation}

To show that the proposed mBIOU metric is a good measure of the fitness of our equivalence class predictions, let us consider its behavior in several thought experiments:
\begin{itemize}
    \item \emph{The predicted sets align perfectly with the ground truth}: mBIOU = 1.0 because the intersections are exactly the same as unions.
    \item \emph{The predicted sets have only one character of overlap with the ground truth, the smallest possible intersection}: mBIOU will be low, because the intersection cardinality will not exceed one, which suppresses IOU for non-trivial equivalence classes.
    \item \emph{The predicted sets contain many false positives for any class, i.e., low precision}: mBIOU would be low due to having relatively large unions.
    \item \emph{The predicted sets contain few true positives for any class, i.e., low recall}: mBIOU would be low due to having relatively small intersections.
\end{itemize}
From these scenarios, we can see that mBIOU is a reasonably good measure of how close our predictions over $U'$ are to ground truth.

It is important to note, however, that $U'$ has some different properties than $U$, which is the entire set of more than 100,000 renderable Unicode codepoints. Critically, assuming that the distribution of cardinality for ground truth equivalence classes remains similar, the number of negatives for each of the classes will be much, much larger. It is therefore crucial to have high precision. To assess the ability of our proposed approach on $U$, we add in new random codepoints to $U'$, and obtain $U''$ and $U'''$ for 1000 and 3000 additions, respectively. 

Importantly, the predicted equivalence classes are measured against $H'_{1}, H'_{2}, \dots, H'_{1313}$, which do not include the new random addition of characters. This is because: 1) we do not know which equivalence classes the new additions belong to; 2) we know that they very likely do not belong to the known equivalence classes on $U'$, otherwise they would have been identified easily by humans who already found the original classes; 3) if we modify our ground truth classes to include these trivial sets of non homoglyphs, then the mBIOU can be inflated if it is good at clustering non-homoglyphs to their own sets, which will not give us a meaningful way of telling whether our model can generalize to $U$ with high precision. We also directly provide a measure of precision, \emph{mean best precision} (mBP), defined as
\begin{gather}
    mBP = \frac{1}{n}\times\sum^{n}_{i=1} \frac{| H'_{i} \cap H^{best}_{i} |}{| H'_{i} \cup H^{best}_{i} |}
\end{gather}
where $H_{best,i}$ is defined in Equation~\ref{eq:best_predicted_class}. Assuming that these random additions of characters do not belong to any of the ground truth equivalence classes on $U'$, for our model to generalize to $U$, we expect mBIOU and mBP to not fall off meaningfully on these random additions of negatives. We also provide mBIOU and mBP for a \emph{naive baseline} approach, where each codepoint belongs to its own predicted equivalence class of size one.

\subsection{Metrics: Comparing to Prior Work}
We also compare our approach to prior work, which focused on identifying homoglyph pairs rather than clusters of equivalence classes. Note that Ginsberg and Yu's work~\cite{ginsberg2018rapid} is an attempt at \emph{rapid} homoglyph detection, and presents little evidence that their approach is also accurate. Sawabe et al.'s OCR approach~\cite{sawabe2019detection} can only detect homoglyphs for ASCII characters, and thus it is not applicable to more than 99\% of Unicode characters. As such we solely focus on Roshanbin and Miller's work~\cite{roshanbin_miller_2011} in our comparison. To this end, we sample pairs of characters from $U'$, which is labeled data. We then calculate the accuracy, precision-recall-curve (PR curve), and average precision for our approach and their approach. To be more specific, let us define all possible 
homoglyph pairs from $U'$ as the set
\begin{gather}
    P = \{(a, b) | a \in U', b \in U', a \neq b, a \equiv b\}
\end{gather}
and all possible 
non-homoglyph pairs from $U'$ as the set
\begin{gather}
    N = \{(a, b) | a \in U', b \in U', a \neq b, a \not\equiv b\}
\end{gather}
In one trial of $n$ pairs (we use $n=2000$), we sample with replacement $n/2$ pairs from $P$ and $n/2$ pairs from $N$. For each pair, we sample characters $a$ and $b$ uniformly at random, where $a \equiv b$ for a $P$ datapoint and $a \not\equiv b$ for an $N$ datapoint. 
Then we obtain the cosine similarity between $a$ and $b$ on our learned embedding of the sampled pair. We also calculate the NCD with the LZMA compressor~\cite{roshanbin_miller_2011}. 
This allows us to train two linear kernel support vector machines, one for the cosine similarities and one for the Normalized Compression Distance, to have large margin classifications for whether a pair belongs to $P$ (positive) or $N$ (negative). We use the compression function in the LZMA library in Python 3 with default parameters. We train the SVMs, calculate the accuracy and average precision, as well as plot the precision-recall curve using the sklearn library.

\begin{figure}[t]
\centerline{\includegraphics[width=.35\textwidth]{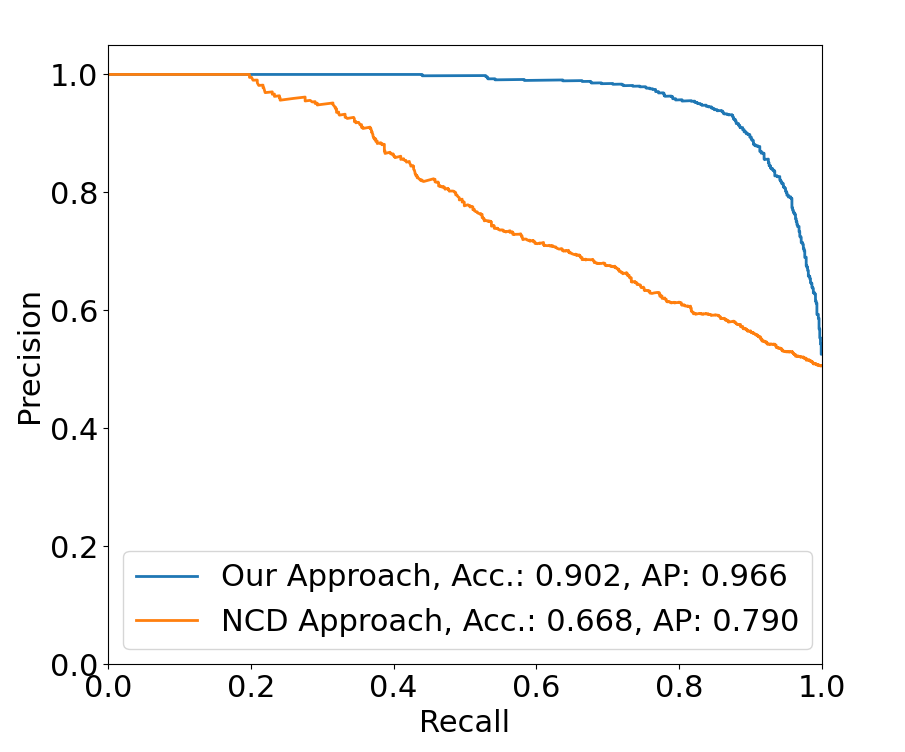}}
\vskip -0.3cm
\caption{PR curve, accuracy (Acc.), and average precision (AP) of our method and NCD on identifying homoglyph and non-homoglyph pairs ($n$ = 2000)}
\label{fig:pr}
\end{figure}

\subsection{Results}
\label{results}
\subsubsection{Clustering Homoglyphs}
Table \ref{table:cluster_metrics} contains the clustering results. It might be confusing to see that the baseline approach achieves precisely the same mBIOU and mBP across all datasets, even the ones augmented with negative examples. This is because the ground truth sets are the same across three datasets, and thus for each $H'_{i}$, we get a $H^{best}_{i}$ that is composed of one character from $H'_{i}$. This results in identical mBIOU and mBP, across all three datasets. Apart from this observation, we see that our approach is better than the baseline by a significant margin. We also see some drop in mBIOU as more negative examples are added, without corresponding drop in mBP, suggesting that the approach needs work on recall before it can generalize well to $U$.

\begin{table}[t]
\renewcommand{\arraystretch}{1.2}
\caption{mBIOU and mBP of Predicted Homoglyph Sets}
\begin{center}
\begin{tabular}{c|ccc}
Data & $U'$ & $U''$ & $U'''$\\
\hline \hline
mBIOU           &0.592&0.580&0.566  \\
mBP             &0.697&0.696&0.696  \\
\hline
mBIOU (Baseline)           &0.430 &0.430 &0.430  \\
mBP (Baseline)             &0.430 &0.430 &0.430  \\
\end{tabular}
\label{table:cluster_metrics}
\end{center}
\vspace{-10pt}
\end{table}
\subsubsection{Comparing to Prior Work}
With an overall accuracy of 0.90 and average precision of 0.97, our method greatly outperforms Roshanbin and Miller's method~\cite{roshanbin_miller_2011} (Figure \ref{fig:pr}) for identifying homoglyph pairs. Given the same number of false positives, our method can potentially identify many more homoglyphs. This also hints that, if applied to clustering homoglyphs into equivalence classes, NCD will likely fare worse than our deep-learning approach. 

\section{Finding Previously Unknown Homoglyphs}
\begin{figure}[t]
\centerline{\includegraphics[width=.40\textwidth]{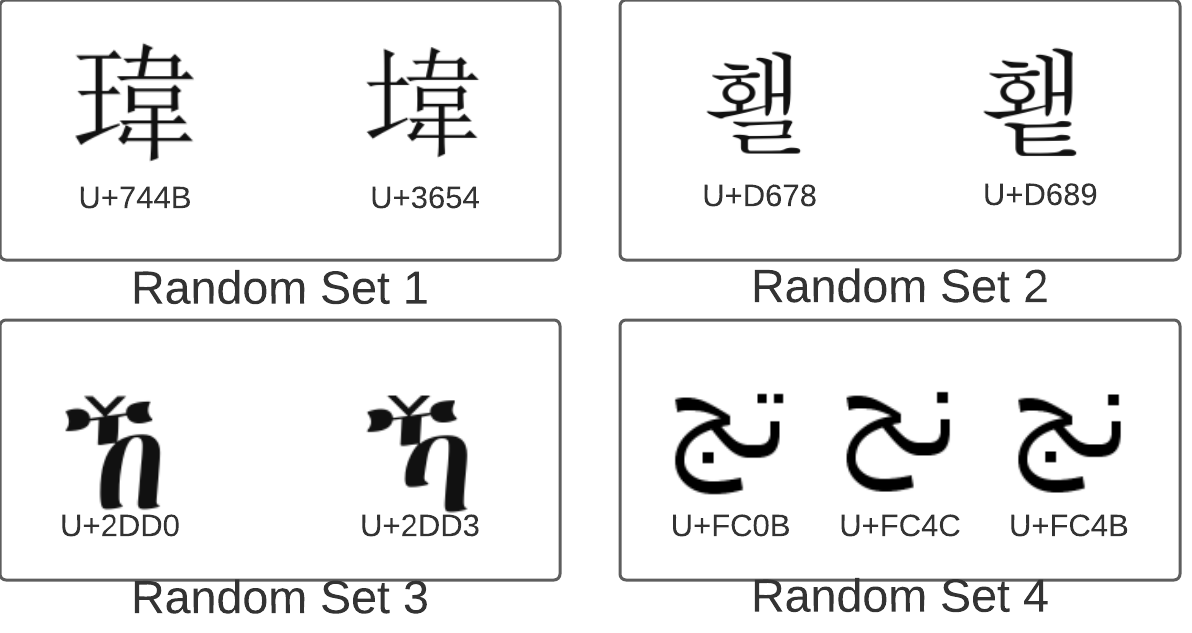}}
\vskip -0.3cm
\caption{Four random sets of predicted homoglyphs previously unidentified}
\vskip -0.5cm
\label{fig:homo_random}
\end{figure}

\begin{figure}[t]
\centerline{\includegraphics[width=.40\textwidth]{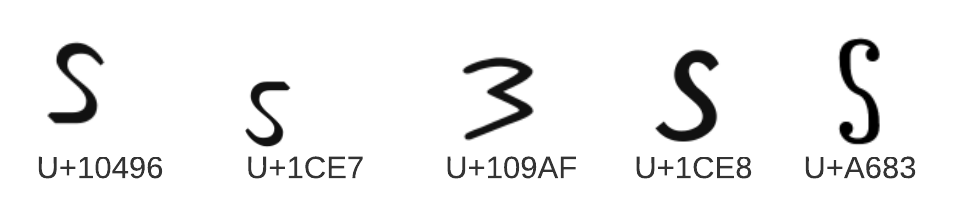}}
\vskip -0.4cm
\caption{A latin-confusable set of predicted homoglyphs previously unidentified}
\label{fig:homo_latin}
\end{figure}

We also test whether our approach can find previously unknown homoglyphs. 
To do this, we utilize the pairwise character cosine similarity matrix $A$ of size $N^2$, where $N=116,294$ is the total number of renderable codepoints. We apply a handpicked threshold $\alpha$ and look for rows in $A$ with more than one element greater than the threshold. Note that each row will always have at least one element passing the threshold due to self-comparison (a cosine similarity of 1.0). We sampled thresholds in the range $-1<\alpha<1$ due to the range of values for cosine similarity. Using an arbitrary value of $\alpha = 0.93$, our model predicts 8452 homoglyphs previously unrecorded by Unicode Consortium, which is \emph{more than two times} the recorded number. We show a random sample of four such sets of predicted homoglyphs below (Figure~\ref{fig:homo_random}). Due to the relevance of Latin Scripts in WWW use today, we also present a set of previously unidentified Latin-confusable homoglyphs (Figure~\ref{fig:homo_latin}). While some of these are false positives, many appear to be genuine homoglyphs. On the other hand, it is arguable whether the characters look visually confusing to native users of the language scripts. The full list of predicted homoglyphs is on Github.

\section{Discussion}
\label{discussion}
Outperforming prior work on homoglyph identification and achieving promising results on clustering Unicode homoglyphs, our work shows that deep learning can effectively generate embeddings for Unicode characters, which can not only be used by attackers to find homoglyphs (as in our case), but also possibly by defenders to detect phishing attacks using homograph strings (see \cite{woodbridge2018detecting} for an example of defense). Further research is still needed, because we did not address composite characters of multiple Unicode codepoints, or composite homoglyphs of multiple characters. Our approach has a far from perfect mBIOU, and can likely be improved by utilizing existing state of the art algorithms that cluster embeddings into an unknown number of classes (e.g., hierarchical clustering). While we predicted 8452 homoglyphs, domain experts or surveys are needed to determine the prediction quality.

\bibliographystyle{IEEEtran}
\bibliography{IEEEabrv,refs}

\end{document}